\documentstyle[12pt]{article}

\newcommand{\e}{{\rm  e}}

\newcommand{\GEV}{ {\rm GeV} }
\newcommand{\TEV}{ {\rm TeV} }

\newcommand{\VEV}[1]{\left\langle #1 \right\rangle}

\begin{document}
\baselineskip 0.7cm

\begin{titlepage}

\begin{center}

\hfill KEK-TH-599\\
\hfill YITP-98-70 \\
\hfill UT-829\\

\vskip 0.5cm
{\large  Large Squark and Slepton Masses 
         for the First-Two Generations \\
         in the Anomalous U(1) SUSY Breaking Models
}
  \vskip 0.5in {\large
    J.~Hisano$^{(a)}$,
    Kiichi~Kurosawa$^{(b)}$, and 
    Yasunori~Nomura$^{(c)}$ }
\vskip 0.4cm 
{\it 
(a) Theory Group, KEK, Oho 1-1, Tsukuba, Ibaraki 305-0801, Japan
}
\\
{\it 
(b) YITP, Kyoto University, Kyoto 606-8502, Japan
}
\\
{\it 
(c) Department of Physics, University of Tokyo, Tokyo 113-0033, Japan
}
\vskip 0.5in

\abstract {Considering that the soft SUSY breaking scalar masses come
 from a vacuum expectation value of the $D$-term for an external gauge
 multiplet, the renormalization of the scalar masses is related to the
 gauge anomaly. 
 Then, if the external gauge symmetry is anomaly-free and has no kinetic 
 mixing with the other U(1) gauge symmetries, the scalar masses are
 non-renormalized at all orders assuming that the gaugino masses
 are negligibly small compared with the scalar masses. 
 Motivated by this, we construct models where the sfermion masses for
 the first-two generations are much heavier than the other
 superparticles in the minimal SUSY standard model in a framework of
 the anomalous U(1) mediated SUSY breaking. 
 In these models we have to introduce extra chiral multiplets with the
 masses as large as those for the first-two generation sfermions.
 We find that phenomenologically desirable patterns for the soft SUSY
 breaking terms can be obtained in the models. 
}
\end{center}
\end{titlepage}
\setcounter{footnote}{0}

Supersymmetry (SUSY) is a solution to the naturalness problem in the
standard model (SM) with the fundamental Higgs boson. 
If the SUSY breaking scale is at most of the order of $100~\GEV$ or
$1~\TEV$, the quadratic divergent radiative correction to the Higgs
boson mass is regularized by the scale. 
On the other hand, introduction of the soft SUSY breaking terms to the
minimal SUSY SM (MSSM) may lead to various flavor problems. 
First, flavor-changing neutral current (FCNC) processes, such as
$K^0$-$\overline{K}^0$ mixing and $\mu\rightarrow e\gamma$, are induced
by the arbitrary squark and slepton masses. 
Second, the CP violating phases in the SUSY breaking parameters may
generate the electric dipole moments (EDMs) of neutron and electron
beyond the experimental bounds. 

It is well-known that these problems can be solved if the first and
second generation sfermions are much heavier than the other
superparticles in the MSSM \cite{1-2_decouple}. 
Also, these masses are irrelevant to the naturalness for the Higgs boson 
masses at one-loop level. 
However, these mass parameters are coupled with each other in the
renormalization group (RG) equations at two-loop level. 
This comes from the $\epsilon$-scalar contribution in the
$\overline{DR}'$ scheme \cite{DR}. 
As a result, if the SUSY breaking masses are generated at high energy
scale, such as the Grand Unified Theory (GUT) scale 
($\sim 10^{16}~\GEV$) or the gravitational scale 
($M_* \sim 10^{18}~\GEV$), the larger soft SUSY breaking masses for the 
first-two generations give a large negative contribution to those for
the third generation, and the SU(3)$_C$ color may be broken
\cite{color_breaking}.
This is a big obstacle to construct models where the first-two generation
sfermions are much heavier than the other superparticles.

The structure of the RG equations for the soft SUSY breaking scalar
masses can be understood by introduction of an external (spurious)
U(1) gauge symmetry \cite{scalar_running}. 
The soft SUSY breaking scalar masses can be regarded as a non-vanishing
$D$-term for the external U(1) gauge multiplet. 
As will be shown later, if the Lagrangian is invariant under the
external U(1) gauge symmetry and if the gauge symmetry is anomaly free
for the SM gauge groups and has no kinetic mixing with them, the SUSY
breaking scalar masses are RG invariant at any orders, assuming that the 
gaugino masses are negligibly small compared with the scalar masses. 
In that case, the dangerous contribution in the RG equations for the
sfermion masses leading to the color breaking ($\epsilon$-scalar
contribution in the $\overline{DR}'$ scheme) automatically vanishes. 

In this letter we show that if the external U(1) gauge symmetry is
anomaly-free, the SUSY breaking scalar masses are non-renormalized at
all orders neglecting the gaugino masses. 
We construct models where the sfermions of the first-two
generations are much heavier than the other superparticles in the MSSM
in a framework of the anomalous U(1)$_X$ mediated SUSY breaking
\cite{anomalous-U1, FCNC-suppress}.
In these models, we have to introduce additional chiral multiplets with
the masses as large as those for the sfermions of the first-two
generations.

First, we review the RG equations for the soft SUSY breaking scalar
masses in SUSY gauge theories \cite{gaugino_running, scalar_running}. 
Here, we use the Wilsonian Lagrangian for simplicity. 
The bare Lagrangian for the SU($N_c$) SUSY gauge theory with the soft
SUSY breaking terms is written as
\begin{eqnarray}
  {\cal L} &=& 
  \left\{ \frac{1}{4} \int d^2\theta~ 
      S {\cal W}^\alpha(V) {\cal W}_\alpha(V) + {\rm h.c.} \right\}
  + \int d^4\theta \Phi_r^{\dagger} \e^{2 V + 2q_r U} \Phi_r,
\end{eqnarray}
introducing a chiral superfield $S$ and a U(1) vector superfield $U$ as
external fields. 
Here, $\Phi_r$ is a chiral multiplet belonging to the $r$ representation
of the SU($N_c$) with the external U(1) charge of $q_r$, and 
${\cal W}^\alpha(V)$ is the field strength for the SU($N_c$) gauge
multiplet $V$ ($V \equiv \sum_{a} {\rm T}_a^{(r)} V_a$).
The generators ${\rm T}_a^{(r)}$ of the SU($N_c$) are normalized as 
$T_r = 1/2$ (${\rm Tr}\, {\rm T}_a^{(r)} {\rm T}_b^{(r)} = 
T_r \delta_{ab}$) in the case of $r$ being the fundamental
representation. 
The vacuum expectation value (VEV) for the scalar component of $S$
corresponds to the gauge coupling constant ($g$) and that for the
$F$-term to the gaugino mass ($m_{\tilde{g}}$) as
\begin{eqnarray}
  \VEV{S} &=& \frac{1}{g^2}(1 -2 m_{\tilde{g}} \theta^2).
\end{eqnarray}
Also, the non-vanishing $D$-term VEV $\VEV{D_U}$ for $U$ leads to the
soft SUSY breaking scalar masses $m_r^2$ for $\Phi_r$, which are
proportional to $q_r$ as\footnote{
The convention for the sign of the $D$-term is as in
Ref.~\cite{convention}.}
\begin{eqnarray}
  m_r^2 &=& q_r \VEV{-D_U}.
\label{soft_mass_squared}
\end{eqnarray}

If the bare Lagrangian is invariant under the external U(1) symmetry and 
the U(1) symmetry is anomaly free for the SU($N_c$) gauge symmetry,
\begin{eqnarray}
  \sum_{r} m_r^2 T_r \propto \sum_{r} q_r T_r =  0,
\label{non-anomalous}
\end{eqnarray}
the scalar masses are RG invariant at all orders due to the
Ward-Takahashi identity assuming that the gaugino mass is negligible. 
In fact, the RG equations for the SUSY breaking scalar masses at
two-loop level in the $\overline{DR}'$ scheme are 
\begin{eqnarray}
  \mu \frac{d m_r^2}{d \mu} &=&
  \frac{8 g^4}{(16\pi^2)^2} C_r \sum_{s} m_s^2 T_s,
\end{eqnarray}
ignoring terms proportional to $m_{\tilde{g}}^2$.
Here, $C_r$ is the Casimir coefficient of the $r$ representation. 
The right-hand side of this equation, that is the $\epsilon$-scalar
contribution, vanishes if relation (\ref{non-anomalous}) is satisfied.

When another U(1)$'$ gauge symmetry exists and $\Phi_r$ has its charge
$q'_r$, the conditions that the kinetic mixing between the internal and
external U(1) gauge symmetries is prevented are
\begin{eqnarray}
  && \sum_{r} m_r^2 q'_r \propto \sum_{r} q_r q'_r =  0, \nonumber \\
  && \sum_{r} m_r^2 q'_r \ln Z_r  \propto \sum_{r} q_r q'_r \ln Z_r =  0.
  \label{non-anomalous-u1}
\end{eqnarray}
Here, $Z_r$ is the wave function renormalization for $\Phi_r$.

From the above discussion, if the soft SUSY breaking scalar masses
satisfy the relations (\ref{non-anomalous}, \ref{non-anomalous-u1}),
the mass spectrum that the first-two generation sfermions are much
heavier than the other superparticles in the MSSM is RG stable. 
To satisfy the relations, however, we have to introduce additional
chiral multiplets which belong to nontrivial representations of the SM
gauge groups with the negative SUSY breaking scalar mass squared. 
In order not to break SU(3)$_C$ or U(1)$_{EM}$, they have to have the
supersymmetric mass terms. 
The supersymmetric masses should be of the same order as the SUSY
breaking masses for the first-two generations so that the
$\epsilon$-scalar contribution is not generated after decoupling of the
additional chiral multiplets.

Now, we construct models which have the above properties.
We adopt the anomalous U(1)$_X$ gauge symmetry, whose anomalies are
canceled by the Green-Schwarz mechanism \cite{GS-mechanism}, as an
origin of the SUSY breaking and identify it as the above external U(1)
symmetry. 
As will be discussed later, non-trivial anomalies for the U(1)$_X$ do
not mean that the U(1)$_X$ gauge symmetry has an anomalous matter
content for the SM gauge groups below the breaking scale of the U(1)$_X$ 
symmetry. 
In the anomalous U(1)$_X$ SUSY breaking models, the Fayet-Iliopoulos
term $\xi$ having mass dimension one is generated of the order of
$O(10^{-(1-2)}) M_*$ \cite{FI-term} and induces the spontaneous SUSY
breaking, so that the $D$-term contribution to the SUSY breaking scalar
masses may dominate over the $F$-term contribution suppressed by $M_*$. 
The correct treatment of the dynamics requires to include the dilaton
field $S$ in the potential \cite{dilaton_global}. 
Even then, depending on the K\"{a}hler potential for the dilaton
field,\footnote{
In general, the K\"{a}hler potential for the dilaton field $S$ can be
written as $K_S(S + S^{\dagger} - \delta_{GS}X)$. 
Here, $X$ is the U(1)$_X$ gauge multiplet and $\delta_{GS}$ is a
constant. 
The $D$-term dominated SUSY breaking occurs when $K_S'' \gg K_S'$, where 
prime denotes derivative with respect to $S$.}
the SUSY breaking can be dominated by the $D$-term and the contribution
from the dilaton field is also negligible \cite{dilaton_SUGRA}. 
We assume that it is the case and neglect the dilaton field $S$,
hereafter.

We choose the anomalous U(1)$_X$ charges for quarks and leptons in the
first-two generations to be $1$ and those for the other chiral
multiplets in the MSSM to be $0$ for simplicity.
The charge assignment that the first-two generations have the same
U(1)$_X$ charges helps to avoid dangerous FCNC processes
\cite{FCNC-suppress}.
The extension to other cases is straightforward. 
As we mentioned above, we have to introduce additional chiral
multiplets, $\psi_{\rm ex}$ and $\bar{\psi}_{\rm ex}$,
with the supersymmetric masses to satisfy the relations
(\ref{non-anomalous}, \ref{non-anomalous-u1}). 
In order not to break the success of the gauge coupling unification, we
introduce $n_5$ pairs of complete SU(5)$_{\rm GUT}$ multiplets in the
fundamental representations,\footnote{
Alternatively, we can also introduce
\begin{eqnarray}
  n_{10} \; [ \psi_{\rm ex}({\bf 10})_{-4/3n_{10}} 
    + \bar{\psi}_{\rm ex}({\bf 10^\star}) _{-4/3n_{10}} ],
\label{extra_10}
\end{eqnarray}
or combinations of Eq.~(\ref{extra_10}) and Eq.~(\ref{extra_5})
maintaining the relations (\ref{non-anomalous},
\ref{non-anomalous-u1}).}
\begin{eqnarray}
  n_5 \; [ \psi_{\rm ex}({\bf 5})_{-4/n_5} 
    + \bar{\psi}_{\rm ex}({\bf 5^\star})_{-4/n_5} ],
\label{extra_5}
\end{eqnarray}
where the numbers in parentheses correspond to representations of the
SU(5)$_{\rm GUT}$ and the subscripts of parentheses denote the U(1)$_X$ 
charges.\footnote{
In fact, this charge assignment satisfies only the one-loop and two-loop 
parts of Eqs.~(\ref{non-anomalous-u1}), that is, $\sum_{r} q_r q'_r =  0$
and $\sum_{r} C_r q_r q'_r =  0$.
Thus, the RG equations for the SUSY breaking scalar masses receive
three-loop contributions of order $O(\alpha_3 \alpha_2 \alpha_Y)$.
These contributions, however, cause no phenomenological problem.}
If we take $n_5 > 4$, the SM gauge couplings blow up below the GUT
scale.

Before constructing explicit models, we review the anomalous U(1)$_X$
mediated SUSY breaking and sketch how to realize the above idea. 
We introduce a SM singlet $\phi$ field with U(1)$_X$ charge $-4/n_5$,
and take the SU($N_c$) gauge theory with one flavor $Q$ and 
$\bar{Q}$ both of which have U(1)$_X$ charges $4/n_5$
\cite{anomalous-U1}.
The tree-level superpotential is given by 
\begin{equation}
  W = \frac{Q \bar{Q}}{M_*}  
      \left( \frac{f_{\phi}}{2} \phi^2 
      + f_{\psi} \psi_{\rm ex} \bar{\psi}_{\rm ex} \right).
\label{sketch_superpotential}
\end{equation}
The dynamical superpotential for $Q$ and $\bar{Q}$ induced by the
SU($N_c$) gauge interaction \cite{exact_sp} leads to 
$\VEV{Q\bar{Q}} \neq 0$. 
Then, the D-flatness condition of the U(1)$_X$ 
\begin{eqnarray}
  \xi^2 - \frac{4}{n_5}|\phi|^2 - \frac{4}{n_5}|\psi_{\rm ex}|^2
    - \frac{4}{n_5}|\bar{\psi}_{\rm ex}|^2 +\cdots
    = 0
\label{D-flatness}
\end{eqnarray}
is not compatible with the $F$-flatness conditions and SUSY is broken.
Moreover, if $f_{\psi}$ is larger than $f_{\phi}$, the VEVs of the 
$\psi_{\rm ex}$ and $\bar{\psi}_{\rm ex}$ fields vanish and the $\phi$
field gets the VEV of order $\xi$, so that the SM gauge groups remain
unbroken.
It can be seen by looking that the scalar potential $V$ behaves as 
\begin{equation}
  V \sim \left| \frac{Q \bar{Q}}{M_*} \right|^2 
    (|f_{\phi}{\phi}|^2 + |f_{\psi}{\psi}_{\rm ex}|^2
    + |f_{\psi}\bar{\psi}_{\rm ex}|^2),
\end{equation}
satisfying the condition Eq.~(\ref{D-flatness}) to the leading order of
$\VEV{Q\bar{Q}}/M_*^2$.

At the minimum of the potential, the $D$-term $D_X$ of the U(1)$_X$ and
the supersymmetric mass $M_{\rm ex}$ for the extra chiral multiplets 
$\psi_{\rm ex}$ and $\bar{\psi}_{\rm ex}$ are written by using the
condensation of $Q\bar{Q}$ as
\begin{eqnarray}
  \VEV{-D_X} = \frac{n_5}{4} 
    \left| \frac{f_{\phi}\VEV{Q\bar{Q}}}{M_*}\right|^2, \qquad
  M_{\rm ex} = f_{\psi}\frac{\VEV{Q\bar{Q}}}{M_*}. \nonumber
\end{eqnarray}
The nonvanishing D-term generates the SUSY breaking scalar masses
according to Eq.~(\ref{soft_mass_squared}).
Note that the squared masses for the scalar components of the 
$\psi_{\rm ex}$ and $\bar{\psi}_{\rm ex}$ fields given by 
$|M_{\rm ex}|^2 -4\VEV{-D_X} /n_5$ are positive under the condition 
$f_{\psi} > f_{\phi}$.
We find that the masses for squarks and sleptons of the first-two
generations are the same order as those of the scalar components of
additional chiral multiplets $\psi_{\rm ex}$ and $\bar{\psi}_{\rm ex}$,
so that the relations (\ref{non-anomalous}, \ref{non-anomalous-u1}) are
satisfied at almost all the energy from high energy scale down to the
weak scale. 
Thus, there are no large radiative corrections to the third generation
sfermions due to the RG equations.

It is essential for the RG stability that supersymmetric masses for the
$\phi$ and $\psi_{\rm ex}$, $\bar{\psi}_{\rm ex}$ fields are comparable, 
that is,
\begin{equation}
  \VEV{\frac{\partial^2 W}{\partial \phi^2}} \sim
    \VEV{\frac{\partial^2 W}
    {\partial \psi_{\rm ex} \partial \bar{\psi}_{\rm ex}}}.
\label{supersymmetric_mass}
\end{equation}
Then, the U(1)$_X$ charge for the $\phi$ field should be equal to those
for the $\psi_{\rm ex}$ and $\bar{\psi}_{\rm ex}$ fields as long as the
couplings $f_{\phi}$ and $f_{\psi}$ are the same order.
Indeed, for the purpose of the SUSY breaking only, we could also
introduce the superpotential
\begin{equation}
  W = \frac{Q\bar{Q}}{M_*} 
      \left( \frac{f_{\phi}}{(n+2)!} \, \frac{\phi^{n+2}}{M_*^{n}} 
      + f_{\psi} \, \psi_{\rm ex} \bar{\psi}_{\rm ex} \right),
\end{equation}
assigning an appropriate U(1)$_X$ charge to the singlet field $\phi$.
In this case, however, the soft SUSY breaking scalar masses 
$\sim \sqrt{-D_X} \simeq \VEV{\partial^2 W/\partial \phi^2}$ are smaller
than the supersymmetric mass $M_{\rm ex} \simeq 
\VEV{\partial^2 W/\partial \psi_{\rm ex} \partial \bar{\psi}_{\rm ex}}$
for the $\psi_{\rm ex}$, $\bar{\psi}_{\rm ex}$ fields by a factor of 
$(\VEV{\phi}/M_*)^n \simeq (\xi/M_*)^n$.
As a result, the relations (\ref{non-anomalous}, \ref{non-anomalous-u1})
do not hold between the two scale $\sqrt{-D_X}$ and $M_{\rm ex}$, giving
large negative contributions to the third generation sfermion squared
masses.
Although it is possible to make phenomenologically viable models in this 
case, we concentrate on the case where Eq.~(\ref{supersymmetric_mass})
holds with $f_{\phi} \simeq f_{\psi}$ and the relations 
(\ref{non-anomalous}, \ref{non-anomalous-u1}) are satisfied at almost
all the energy.

In order to generate the complete Yukawa matrices, we have to introduce
a field with U(1)$_X$ charge $-1$ which has a VEV of the order of $\xi$, 
since otherwise the mixing between the first-two generations with the
U(1)$_X$ charge $1$ and the third generation with the U(1)$_X$ charge
$0$ is not allowed.
In the above simplified model, however, the $\phi$ field has a U(1)$_X$
charge $-4/n_5$, so that we have to choose $n_5 = 4$ resulting in the
asymptotic nonfreedom of the SM gauge couplings.
Thus, we instead introduce two singlet fields $\phi_1$ and
$\phi_2$ which have U(1)$_X$ charges $-1$ and $1-8/n_5$, respectively.

We, here, construct explicit models taking $n_5 = 2$ for
simplicity.\footnote{
When $n_5=2$, the charge assignment of the U(1)$_X$ for the first-two
generations and the additional chiral multiplets is the same as that of
a U(1) in E$_6$ for {\bf 27}.}
We introduce $N_f$ flavors $Q^i$ and $\bar{Q}_{\bar{\imath}}$ 
$(i,\bar{\imath} = 1,\cdots,N_f)$ of the SU($N_c$) gauge theory 
($N_f < N_c$).
The tree-level superpotential is
\begin{equation}
  W = \frac{Q^i \bar{Q}_{\bar{\imath}}}{M_*} 
      \left( f_{\phi}{}^{\bar{\imath}}_i \, \phi_1 \phi_2 
      + f_{\psi}{}^{\bar{\imath}}_i \, \psi_{\rm ex} 
      \bar{\psi}_{\rm ex} \right),
\end{equation}
and the dynamical superpotential is generated as \cite{exact_sp}
\begin{equation}
  W_{\rm dyn} = (N_c-N_f)\left(\frac{\Lambda^{3N_c-N_f}}
                {Q\bar{Q}}\right)^{\frac{1}{N_c-N_f}},
\end{equation}
where $\Lambda$ is the dynamical scale of the SU($N_c$) gauge
theory.
The dynamical degrees of freedom can be expressed by meson fields
\begin{equation}
  M^i_{\bar{\imath}} = Q^i \bar{Q}_{\bar{\imath}}.
\end{equation}
Then, the full superpotential is expressed as
\begin{equation}
  W = \frac{1}{M_*} \phi_1 \phi_2 {\rm Tr}(f_{\phi} M)
      + (N_c-N_f)\left(\frac{\Lambda^{3N_c-N_f}}
                {\det M}\right)^{\frac{1}{N_c-N_f}}.
\end{equation}
Here, we have ignored the $\psi_{\rm ex}$ and $\bar{\psi}_{\rm ex}$ 
fields which have vanishing VEVs provided that $f_{\psi}$ is
sufficiently larger than $f_{\phi}$.
The $D$-term is
\begin{eqnarray}
  -D_X = g_X^2 \left( \xi^2 + 4{\rm Tr}\sqrt{M^{\dagger}M} 
               - |\phi_1|^2 - 3|\phi_2|^2 \right),
\end{eqnarray}
along the SU($N_c$) flat directions.

The minimum of the potential can be obtained by making an expansion in
the parameter $(\Lambda/M_*)^{(3N_c-N_f)/N_c} \ll 1$.
To the leading order, the VEVs of the fields $\phi_1$, $\phi_2$, and $M$
at the minimum are given by solving the following equations:
\begin{eqnarray}
  |\phi_1|^2 + 3 |\phi_2|^2 &=& \xi^2, 
  \label{phi_1} \\
  |\phi_1|^2 \left( (N_f-N_c)|\phi_1|^2 + N_f|\phi_2|^2 \right) &=& 
    3 |\phi_2|^2 \left( (N_f-N_c)|\phi_2|^2 + N_f|\phi_1|^2 \right), 
  \label{phi_2} \\
  (M^{-1}){}^{\bar{\imath}}_i 
    \left(\frac{\Lambda^{3N_c-N_f}}{\det M}\right)^{\frac{1}{N_c-N_f}} &=& 
    \frac{f_{\phi}{}^{\bar{\imath}}_i}{M_*} \phi_1 \phi_2.
  \label{M}
\end{eqnarray}
From Eqs.~(\ref{phi_1}, \ref{phi_2}), we find that both $\phi_1$ and
$\phi_2$ fields have nonvanishing VEVs of the order $\xi$, and the VEV of
the $M$ field is given by Eq.~(\ref{M}) as
\begin{eqnarray}
  \VEV{\phi_1} \simeq \VEV{\phi_2} \simeq \xi, \qquad
  \VEV{M} \simeq \Lambda^{\frac{3N_c-N_f}{N_c}} 
    \left( \frac{M_*}{\xi^2} \right)^{\frac{N_c-N_f}{N_c}}.
\label{VEVap}
\end{eqnarray}
Then, the complete Yukawa matrices are generated through the VEV of
the $\phi_1$ field suppressed by the suitable power of
$\VEV{\phi_1}/M_* \simeq \xi/M_*$ \cite{mass_hierarchy}. 
Since we have chosen the Higgs bosons to be neutral under the U(1)$_X$,
this explains why the third generation quarks and leptons are much
heavier than the first-two generation ones
\cite{anomalous_FN}.\footnote{
If we assign the different charges among the first-two generation quarks 
and leptons, it may be possible to obtain more realistic mass matrices.}

The auxiliary fields are determined as
\begin{eqnarray}
  -{F}_{\phi_1}^{\dagger} &=& \frac{1}{M_*} \phi_2 
     {\rm Tr}(f_{\phi} M), 
   \label{fterm1}\\
  -{F}_{\phi_2}^{\dagger} &=& \frac{1}{M_*} \phi_1 
     {\rm Tr}(f_{\phi} M), 
   \label{fterm2}\\
  -D_X &=& \frac{2N_f-N_c}{N_c}\frac{1}{\xi^2 M_*^2}
     (|\phi_1|^2 + |\phi_2|^2) \left| {\rm Tr}(f_{\phi} M) \right|^2.
  \label{D-term}
\end{eqnarray}
From Eq.~(\ref{D-term}), $2N_f$ must be larger than $N_c$ in
order to give the positive SUSY breaking squared masses for the
first-two generation sfermions. 
Inserting Eqs.~(\ref{VEVap}) into 
Eqs.~(\ref{fterm1}, \ref{fterm2}, \ref{D-term}), we get 
\begin{eqnarray}
  -{F}_{\phi_1}^{\dagger} \simeq -{F}_{\phi_2}^{\dagger}
    \simeq \left( \frac{\Lambda^{3N_c-N_f}}
    {\xi^{2(N_c-N_f)}M_*^{N_f}} \right)^{\frac{1}{N_c}} \xi, \qquad
  -D_X \simeq \left( \frac{\Lambda^{3N_c-N_f}}
    {\xi^{2(N_c-N_f)}M_*^{N_f}} \right)^{\frac{2}{N_c}}.
\end{eqnarray}
Note that there are the following useful relations among the $D$-term
and the $F$-terms:
\begin{eqnarray}
  -D_X \simeq 
    \left| \frac{-{F}_{\phi_1}}{\VEV{\phi_1}} \right|^2 \simeq 
    \left| \frac{-{F}_{\phi_2}}{\VEV{\phi_2}} \right|^2.
\label{useful_relation}
\end{eqnarray}

So far, we have considered that the U(1)$_X$ have an anomaly-free matter 
content for the SM gauge groups to satisfy the relation
(\ref{non-anomalous}).
However, the U(1)$_X$ has to have anomalies for any gauge groups, so
that there must be chiral multiplets $\chi$ and $\bar{\chi}$ which carry
the U(1)$_X$ anomalies for the SM gauge groups. 
Since the U(1)$_X$ is broken, these chiral multiplets can get the
supersymmetric masses of the order $\xi$ through the coupling with the
$\phi$ ($\phi_1$ and/or $\phi_2$) fields as
\begin{eqnarray}
  W \sim \phi \chi \bar{\chi},
\end{eqnarray}
provided that these fields are vector-like under the SM gauge groups and
have appropriate U(1)$_X$ charges.\footnote{
It may be inevitable that the squared masses for the third generation
sfermions get negative contributions through the RG equations above the
decoupling scale of $\chi$ and $\bar{\chi}$.
However, we can make these negative contributions smaller than the
positive contributions induced by the gravity or gauge mediation.}
Then, the gaugino masses are induced at one-loop level as
\cite{gauge_mediation} 
\begin{eqnarray}
  m_{\tilde{g}} \simeq \frac{\alpha}{4\pi} 
    \left( \frac{-\bar{F}_{\phi}}{\VEV{\phi}} \right),
\end{eqnarray}
where $\alpha$ denotes the SM gauge coupling strength.
The sfermion masses of the same order are also generated at two-loop
level.
This seems the hybrid scenario with the gauge-mediated SUSY breaking.
However, the existence of the messenger fields $\chi$ and $\bar{\chi}$
is not the extra assumption but the natural consequence of the anomalous 
U(1)$_X$ gauge symmetry \cite{hybrid}.

We finally discuss the soft SUSY breaking parameters. 
In the present models, there are three contributions, that is, U(1)$_X$
$D$-term induced $m_D$, gravity mediated $m_F$, and gauge mediated $m_G$
pieces,
\begin{eqnarray}
\begin{array}{ccccccc}
  m_D &\simeq& \sqrt{-D_X}, &&&& \\ 
  m_F &\simeq& {\displaystyle \frac{-\bar{F}_{\phi}}{M_*}}
    &\simeq& {\displaystyle \frac{\xi}{M_*} \left( 
    \frac{-\bar{F}_{\phi}}{\VEV{\phi}} \right)}
    &\simeq& {\displaystyle \frac{\xi}{M_*} m_D}, \\ 
  m_G &\simeq& {\displaystyle \frac{\alpha}{4\pi} 
    \left( \frac{-\bar{F}_{\phi}}{\VEV{\phi}} \right)}
    &\simeq& {\displaystyle \frac{\alpha}{4\pi} m_D.} &&
\end{array}
\label{various_contribution}
\end{eqnarray}
Here, we have used Eq.~(\ref{useful_relation}).
The SUSY breaking scalar masses for the first-two generations
$m_{1,2}$, those for the third generation or Higgs bosons $m_{3,h}$,
gaugino masses $m_{\tilde{g}}$, and analytic trilinear couplings $A$ for 
the sfermions are given by
\begin{eqnarray}
\begin{array}{ccccccc}
  m_{1,2}^2     &\simeq& m_D^2 &+& m_F^2 &+& m_G^2, \\
  m_{3,h}^2     &\simeq&       & & m_F^2 &+& m_G^2, \\
  m_{\tilde{g}} &\simeq&       & &       & & m_G,   \\
  A             &\simeq&       & & m_F. & & 
\end{array}
\end{eqnarray}
From Eq.~(\ref{various_contribution}), we find that the relation 
$m_D \gg m_F \simeq m_G$ holds in a generic parameter region.
In that case, we obtain phenomenologically desirable spectrum at the
low energy, which solves the FCNC and CP problems naturally by making
the first-two generation sfermions much heavier than the other
superparticles in the MSSM.
The point is that this mass spectrum is RG stable, since 
relations (\ref{non-anomalous}, \ref{non-anomalous-u1}) are satisfied
due to the extra chiral multiplets $\psi_{\rm ex}$ and 
$\bar{\psi}_{\rm ex}$.

In conclusion, we have considered that the soft SUSY breaking scalar
masses come from a vacuum expectation value of the $D$-term for an
external U(1) gauge multiplet.
Then, if the external gauge symmetry is anomaly-free and has no kinetic
mixing with the other U(1) gauge symmetries, the scalar masses are
non-renormalized at all orders, assuming that the gaugino masses are
negligibly small compared with the scalar masses. 
Motivated by this, we have constructed the anomalous U(1)$_X$ mediated
SUSY breaking models.
In a generic parameter region of the models, the sfermion masses for the 
first-two generations are much heavier than the other superparticles in
the MSSM, which solves various flavor problems such as FCNC and CP ones.
We hope that the present models give a new possibility in constructing
the models which have such a hierarchical mass pattern of
superparticles.

{\bf Acknowledgments}

We would like to thank N. Haba and H. Nakano for useful discussions.
This work is partially supported in part by Grants-in-aid for Science
and Culture Research for Monbusho, No.10740133 and "Priority Area:
Supersymmetry and Unified Theory of Elementary Particles (\#707)".
Y.N. is supported by the Japan Society for the Promotion of Science.

%
%%%%%%%%%%%%%%%%%%%%%%%%%%%%%%%%%%%%%%%%%%%%%%%%%%%%%%%%%%%%%%%
%
% NEW COMMANDS FOR THE BIBLIOGRAPHY
%
%%%%%%%%%%%%%%%%%%%%%%%%%%%%%%%%%%%%%%%%%%%%%%%%%%%%%%%%%%%%%%%
\newcommand{\Journal}[4]{{\sl #1} {\bf #2} {(#3)} {#4}}
\newcommand{\APJ}{Ap. J.}
\newcommand{\CJP}{Can. J. Phys.}
\newcommand{\NC}{Nuovo Cim.}
\newcommand{\NP}{Nucl. Phys.}
\newcommand{\PL}{Phys. Lett.}
\newcommand{\PR}{Phys. Rev.}
\newcommand{\PRep}{Phys. Rep.}
\newcommand{\PRL}{Phys. Rev. Lett.}
\newcommand{\PTP}{Prog. Theor. Phys.}
\newcommand{\SJNP}{Sov. J. Nucl. Phys.}
\newcommand{\ZP}{Z. Phys.}
%%%%%%%%%%%%%%%%%%%%%%%%%%%%%%%%%%%%%%%%%%%%%%%%%%%%%%%%%%%%%%%

%
\end{document}